\documentclass[preprint2]{aastex}

\usepackage{graphicx}
\usepackage{lscape}
\shorttitle{Orbital-phase resolved spectroscopy of FO Aqr}
\shortauthors{Pek\"on Y. and Balman \c{S}.}

\begin{document}
\title{Orbital phase-resolved spectroscopy of the intermediate polar FO Aqr using \emph{XMM-Newton} Observatory
data}

\author{Y. Pek\"on\altaffilmark{1,2} and \c{S}. Balman\altaffilmark{1,3}}

\altaffiltext{1} {Middle East Technical University, Physics
Department, Inonu Bulvari,
06531, Ankara, Turkiye}
\altaffiltext{2} {yakup@astroa.physics.metu.edu.tr}
\altaffiltext{3}{solen@astroa.physics.metu.edu.tr}

\begin{abstract}

We present the orbital-phase resolved analysis of an archival FO Aqr
observation obtained using the X-ray Multi-Mirror Mission
(\emph{XMM-Newton}), European Photon Imaging Camera (pn instrument).
We investigate the variation of the spin pulse amplitudes over the
orbital period in order to account for the effects of orbital motion
on spin modulation. The semi-amplitude variations are in phase with
the orbital modulation, changing from (38.0$\pm$1.8)\% at the
orbital maximum to (13.3$\pm$3.7)\% at the orbital minimum. The
spectral parameters also show changes over the orbital period. One
of the absorption components increase by a factor of 5 between the
orbital minimum and maximum. We interpret that this absorption
arises from the bulge where accretion stream from the secondary
impacts the disk. The spectrum extracted from the orbital minima and
maxima can be fitted with a warm absorber model yielding values
$N_{\rm{H}}$ = 2.09$^{+0.98}_{-1.09}$$\times$ $10^{22}$ and
0.56$^{+0.26}_{-0.15}$$\times$ $10^{22}$ cm$^{-2}$; and log($\xi$) =
0.23$^{+0.37}_{-0.26}$ and $<$0.30 erg cm s$^{-1}$ respectively,
indicating the existence of ionized absorption from the bulge at the
impact zone which is spread out on the disk. The absorption due to
accretion curtain and/or column which causes the spin modulation can
be distinguished from the disk absorption via spectral modeling.

\end{abstract}


\section{Introduction}

FO Aqr is a compact binary belonging to the sub-category
intermediate polars (IPs) of the cataclysmic variables (CVs) which
are composed of a white dwarf accreting material from a Roche lobe
filling main sequence star. IPs have white dwarfs with a magnetic
field of mostly 1-10 MG (although there are a couple of systems with
magnetic field strengths up to 30 MG; e.g. Katajainen et al. 2010,
Piirola et al. 2008). The accretion occurs through a truncated disk
and via accretion curtains to the magnetic poles of the white dwarf
(see Patterson 1994; Warner 2003).

FO Aqr is a well studied IP, with an orbital period of 4.85 hr
(Osborne \& Mukai 1989) and white dwarf spin period of 20.9 min
(Patterson et al. 1998). It has been observed with almost every
X-ray mission; \emph{EXOSAT}, \emph{Ginga}, \emph{ASCA},
\emph{RXTE}, \emph{XMM-Newton}, \emph{INTEGRAL}, \emph{SWIFT} and
\emph{Suzaku} (Cook, Watson \& McHardy 1984, Norton et al. 1992;
Mukai, Ishida \& Osborne 1994; Evans et al. 2004, Parker, Norton \&
Mukai 2005, Landi et al. 2009, Yuasa et al. 2010, respectively). Our
understanding of the accretion scenario in FO Aqr has changed over
the years: Norton et al. (1992) suggested that the system shows
diskless accretion and that the accretion flow changes poles.
However, Hellier (1993) argued that there is an accretion disk in
the system and the accretion takes place both via a disk and stream
overflowing the disk. Mukai, Ishida \& Osborne (1994) confirmed this
hybrid accretion mode, but proposed that accretion from a partial
disk is dominant. Later, it was suggested that the accretion mode
alternates from a hybrid of disk-fed and stream fed accretion to
disk-fed accretion over the years (i.e. hybrid in 1988, disk-fed in
1990 (Norton, Beardmore \& Taylor 1996) hybrid in 1993 and 1998
(Beardmore et al. 1998), disk-fed in 2001 (Evans et al. 2004)). The
changes in accretion modes are most likely due to changes in the
mass accretion rate of the system (de Martino et al. 1999). The
source shows orbital modulations, which are deeper at the lower
energy regime, indicating that it could be due to absorption from
structures on the disk and the accretion stream (Hellier et al.
1993; Evans et al. 2004; Parker, Norton \& Mukai 2005). The spin
pulse shape of the system is complicated, with a quasi-sinusoiadal
component and a notch after a "dip" caused by the accretion curtains
(Evans et al. 2004). The spin pulse profile may vary between nearly
sinusodial to saw-tooth shapes in different observations (Beardmore
et al. 1998).

The X-ray spectrum of the source can be represented with a multiple
plasma emission component, complex absorption and Gaussian lines
(eg. Mukai, Ishida \& Osborne 1994; Evans et al. 2004; Yuasa et al.
2010). A soft blackbody emission was not detected by Evans \&
Hellier (2007) with \emph{XMM-Newton} data or by Yuasa et al. (2010)
with \emph{Suzaku} data. However, using the results of a combined
joint analysis of \emph{INTEGRAL/IBIS} and \emph{SWIFT/XRT} data,
Landi et al. (2009) claimed the detection of a blackbody component
with a temperature of 61 eV which they attributed to the irradiation
of the white dwarf atmosphere.

In this work, the spectral and temporal properties of FO Aqr over
the orbital phase will be investigated. In Section 2 observation and
data preparation will be introduced. In Section 3 the spin pulse
profile behavior over the orbital phase will be outlined. In Section
4 the phase average spectrum and the variation of spectral
parameters over the orbital phase will be investigated. In Section 5
the results of the analyses will be discussed.

\section{Observation and Data Analysis}

FO Aqr was observed for about 35 ks (exposure time) on 12 May 2001
(OBS ID: 0009650201) with \emph{XMM-Newton} Observatory (Jansen et
al. 2001) with all of its instruments on board: a pn CCD detector
(Str\"{u}der et al. 2001), two MOS CCD detectors (Turner et al.
2001) sensitive in the 0.2-15 keV energy range at the focus of three
European Photon Imaging Cameras (EPIC), Reflecting Grating
Spectrometers (RGS; den Herder et al. 2001), a high-resolution
spectrometer working together with the EPIC detectors, and the
optical monitor (OM) instrument with an optical/UV camera (Mason et
al. 2001).

Due to the higher sensitivity compared with the MOS, we used the
archival EPIC pn data in the analysis, obtained using the small
window mode, with a net count rate of 2.33 $\pm$ 0.01 c s$^{-1}$ in
the 0.3-10.0 keV interval. Standard data preparation and processing
was carried out with the raw data using the \emph{XMM-Newton}
Science Analysis Software ({\sc sas}), version 11.0.0. The
extraction of the background subtracted source spectrum was carried
out using {\sc especget} and the extraction of the source and
background light curve was carried out with {\sc evselect} which are
tools within {\sc sas}. Data with single- and double-pixel events,
i.e. patterns 0-4 with Flag=0 options were extracted at all times. A
circular extraction region with a radius of 32.4 arcsec was used for
the source and background photons. To extract the light curve and
spectra resolved at the orbital phase, we used the {\sc phasecalc}
tool to create a phase column in the events file with observation
start time being the 0 phase. Next, phase-resolved spectra were
extracted from the modified events file using a script. Further
analyses of light curve and spectra were performed with {\sc xronos
5.2} and {\sc xspec 12.6.0} packages. We obtained the background
subtracted source light curve using {\sc lcmath} tool in {\sc
ftools} v6.10 and applied barycentric correction. Our preliminary
analysis of FO Aqr is summarized in Balman \& Pek\"on (2011). The
original analysis of the archival \emph{XMM-Newton} data of FO Aqr
can be found  in Evans et al. (2004).

\section{Spin Modulation Variations over the Orbit}

We folded the corrected light curve over the orbital (17458 s) and
spin (1254 s) periods (Patterson et al. 1998) using the same epoch
given by Evans et al. (2004); T$_0$=2452041.806 (HJD) for the
orbital period and T$_0$=2452041.858 (HJD) for the spin period.
Figure 1 shows the folded light curve over the orbital period and
the spin period separately (see also, Evans et al. 2004).

\begin{figure}[ht!]
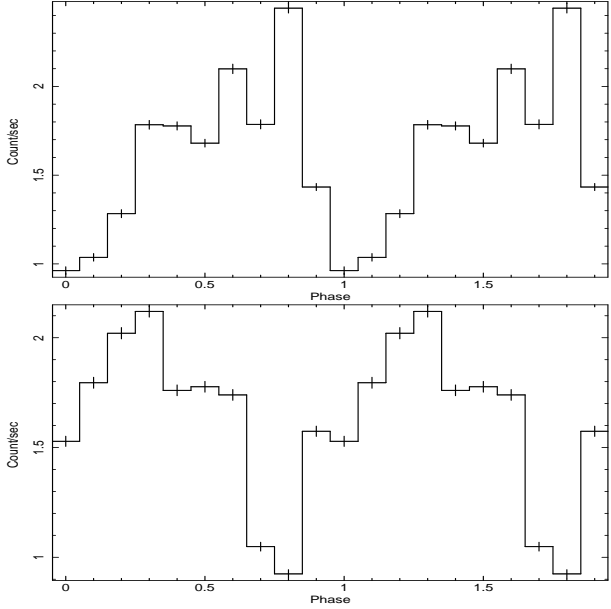

\centerline{
\includegraphics[width=4cm, height=8cm, angle=270]{orbital_evanseph.ps}}
\centerline{
\includegraphics[width=4cm, height=8cm,angle=270]{spin_evanseph.ps}}

\caption{The light curve folded over the orbital period of 17458 s
on top, and the light curve folded over the spin period of 1254 s at
the bottom.}
\end{figure}

In order to investigate the changes in the spin pulse over the
orbital phase, we extracted light curves for each 0.1 orbital phase
interval in the 0.3-10.0 keV range and then folded these over the
spin period. One orbital-phase bin is 1745.8 s, which is greater
than the spin period. In general, the pulse shapes look similar with
spin minima at around phase 0.8. However, we find that the amplitude
of the spin pulses change with orbital phase (See Figure 2). To
construct Figure 2, we fitted spin pulse with a simple sine curve
for each orbital phase. Then we calculated the percentage variations
by taking the ratio of the semi-amplitude of each fit to the mean
value and plotted the percentage variations vs. orbital phase. We
also compared them with the orbital folded light curve. As seen in
Figure 3, the percentage variation is clearly in phase with the
orbital motion. The variations change from (38.0$\pm$1.8)\% at the
orbital maximum to (13.3$\pm$3.7)\%
at the orbital minimum.

\begin{figure}

\includegraphics[width=6cm, height=22cm]{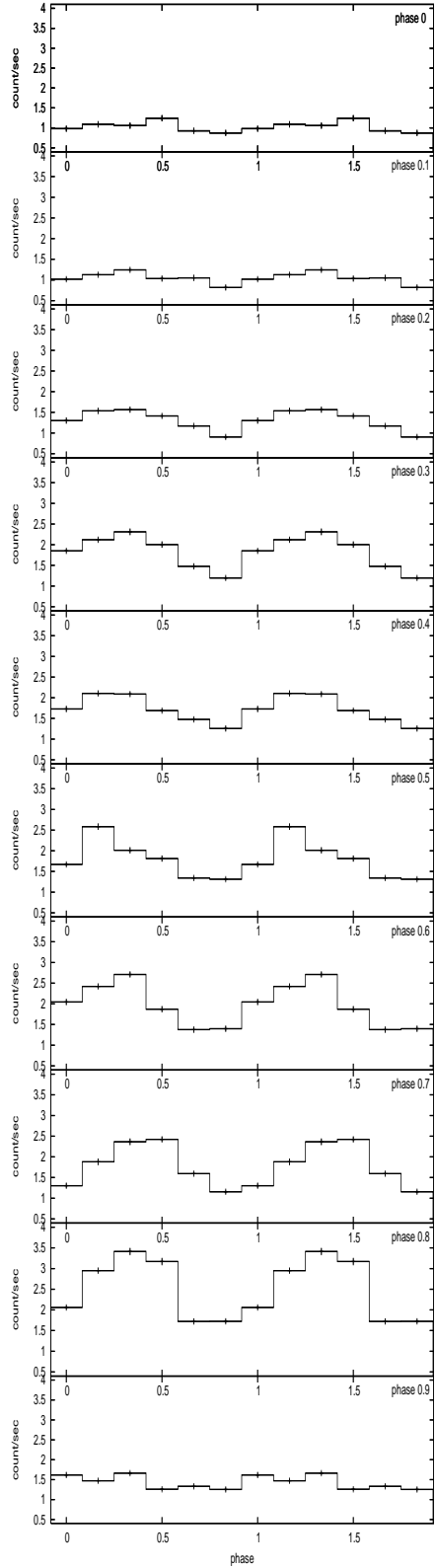}

\caption{The spin pulse profile of FO Aqr for each 0.1 orbital phase
interval. The corresponding phases are noted on each panel.}
\end{figure}

\section{Spectral Variations over the Orbit}

\begin{figure*}[ht!]
\centerline{
\includegraphics[width=8cm,height=16cm,angle=270]{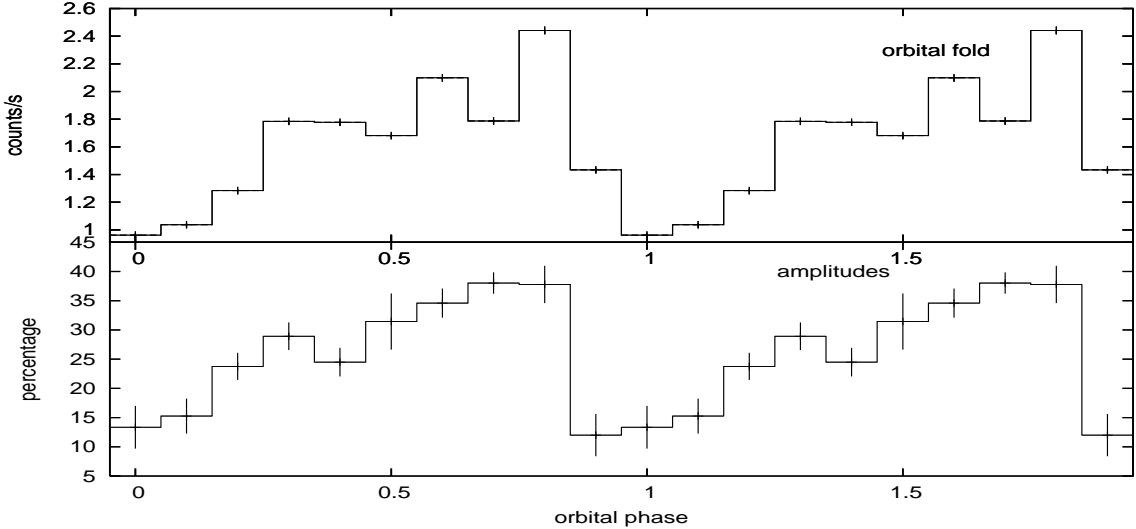}}

\caption{The percentage variation of the spin pulse amplitudes from the mean (below),
together with the orbital flux variation (above).}
\end{figure*}

We used the same composite spectral model from
Evans et al. (2004) in order to
fit the phase average spectrum of the source. The composite model
consists of a simple absorption model ({\sc wabs}), two partial
covering absorption models ({\sc pcfabs}); three plasma emission
models at different temperatures ({\sc mekal}) and a Gaussian
emission line at 6.4 keV (Fe K$_{\alpha}$).

\begin{figure*}[ht!]
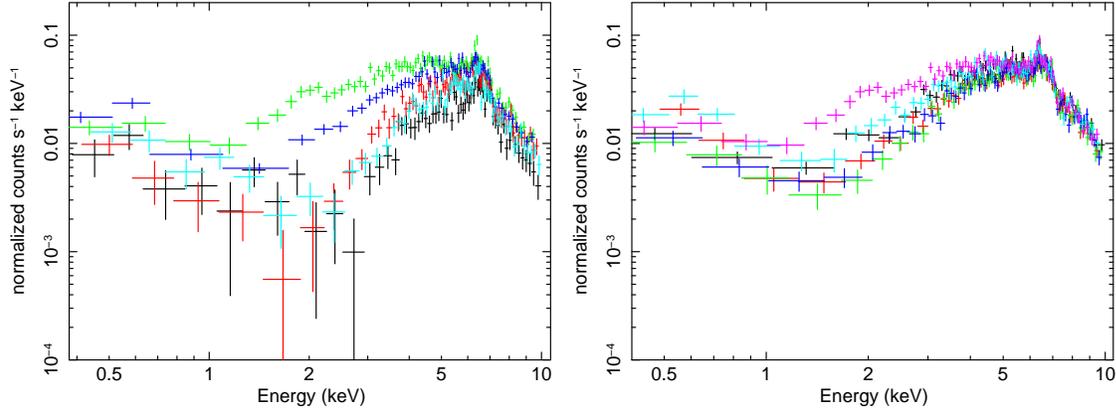

\centerline{
\includegraphics[scale=0.3,angle=270]{fo_orbit_sp2_up.ps}
\includegraphics[scale=0.3,angle=270]{fo_orbit_sp1_up.ps}}
\caption{The orbital-phase resolved spectra of FO Aqr for each 0.1 phase bin interval. The left-hand panel shows phases 0.8 (green)
0.9 (dark-blue), 0.0 (red), 0.1 (black) and 0.2 (light-blue). The right-hand panel shows the phases 0.8 (purple-pink) and the
rest of the phases 0.3-0.7. Notice the spectral differences on the left-hand side during orbital minima and how the spectra are different
for the rest of the phases.}
\end{figure*}

\begin{table*}[ht!]
\centering
%\begin{minipage}{140mm}
\label{1} \scriptsize{ \caption{Spectral parameters derived from the
fits to the spectra for each orbital phase of 0.1 in the 0.3-10 keV
range. All the spectra are fitted with a composite model of three
collisional equilibrium plasma emission models at different
temperatures ({\sc mekal}), 1 Gaussian centered at 6.4 keV, 2
partial covering absorber model for intrinsic absorption ({\sc
pcfabs}) and a simple absorber for the interstellar absorption ({\sc
wabs}). The given errors correspond to 90\% confidence level for a
single parameter.} }

%\begin{center}
\begin{tabular}{@{}llrrrrrrrrrrr@{}}
\hline \hline

\multicolumn{1}{l}{Model} & \multicolumn{1}{l}{Component}&
\multicolumn{1}{r}{0.1} & \multicolumn{1}{r}{0.2} &
\multicolumn{1}{r}{0.3} & \multicolumn{1}{r}{0.4} &
\multicolumn{1}{r}{0.5} \\

\hline

wabs & $N_{\rm{H}}$ ($\times$ $10^{22}$ cm$^{-2}$) &
0.152$^{+0.041}_{-0.029}$ & 0.093$^{+0.048}_{-0.034}$ &
0.077$^{+0.027}_{-0.035}$ & 0.067$^{+0.034}_{-0.027}$ &
0.188$^{+0.024}_{-0.020}$ \\

pcfabs1 & $N_{\rm{H}}$ ($\times$ $10^{22}$ cm$^{-2}$) &
20.0$^{+2.6}_{-2.0}$ & 21.2$^{+1.4}_{-1.3}$ &
18.4$^{+1.1}_{-1.0}$ & 24.6$^{+1.2}_{-1.1}$ & 23.8$^{+1.7}_{-1.5}$
\\ [1ex]

& CoverFrac & 0.689$^{+0.041}_{-0.041}$ & 0.820$^{+0.020}_{-0.020}$ & 0.820$^{+0.015}_{-0.015}$ &
0.836$^{+0.011}_{-0.011}$ & 0.651$^{+0.015}_{-0.015}$
\\ [1ex]

pcfabs2 & $N_{\rm{H}}$ ($\times$ $10^{22}$ cm$^{-2}$)
& 17.45$^{+1.20}_{-1.11}$ & 9.79$^{+0.68}_{-0.63}$ &
7.12$^{+0.44}_{-0.40}$ & 6.37$^{+0.37}_{-0.34}$ &
7.06$^{+0.29}_{-0.27}$
\\ [1ex]

& CoverFrac & 0.980$^{+0.004}_{-0.005}$ & 0.984$^{+0.004}_{-0.007}$ & 0.980$^{+0.003}_{-0.005}$ &
0.980$^{+0.003}_{-0.004}$ & 0.980$^{+0.002}_{-0.002}$
\\ [1ex]

MEKAL1 & kT & 0.103$^{+0.009}_{-0.011}$ & 0.117$^{+0.013}_{-0.015}$ & 0.159$^{+0.025}_{-0.027}$ &
0.165$^{+0.029}_{-0.030}$ & 0.125$^{+0.010}_{-0.010}$  \\

& Norm ($\times$ $10^{-3}$) & 5.36$^{+2.07}_{-2.07}$ & 8.05$^{+3.23}_{-3.23}$ &
3.95$^{+1.43}_{-1.43}$ & 3.61$^{+1.36}_{-1.36}$ & 7.12$^{+1.70}_{-1.70}$ \\
[1ex]

MEKAL2 & kT & 2.17$^{+0.57}_{-0.50}$ & 6.87$^{+5.00}_{-4.17}$ & 2.25$^{+0.57}_{-0.51}$ &
2.71$^{+0.71}_{-0.57}$ & 8.19$^{+3.51}_{-3.00}$  \\

& Norm ($\times$ $10^{-5}$) & 1.90$^{+0.96}_{-0.96}$ & 1.72$^{+0.84}_{-0.84}$ &
2.63$^{+1.25}_{-1.25}$ & 3.01$^{+1.01}_{-1.01}$ & 2.49$^{+0.86}_{-0.86}$  \\
[1ex]

MEKAL3 & kT & 69.1$^{+15.0}_{-15.0}$ & 44.7$^{+18.4}_{-17.6}$ & 27.2$^{+10.2}_{-7.0}$ &
21.3$^{+6.9}_{-3.6}$ & 37.6$^{+9.9}_{-8.4}$ \\

& Norm ($\times$ $10^{-3}$) & 3.57$^{+0.15}_{-0.15}$ & 5.62$^{+0.17}_{-0.17}$ & 5.70$^{+0.15}_{-0.15}$ &
6.85$^{+0.17}_{-0.17}$ & 6.61$^{+0.14}_{-0.14}$ \\

Gaussian & line centre (keV)
& 6.39$^{+0.03}_{-0.03}$ & 6.38$^{+0.04}_{-0.03}$ &
6.41$^{+0.03}_{-0.03}$ & 6.37$^{+0.05}_{-0.04}$ &
6.39$^{+0.02}_{-0.03}$  \\

& $\sigma$
& 0.070$^{+0.069}_{-0.070}$ & 0.052$^{+0.073}_{-0.052}$ &
0.040$^{+0.057}_{-0.040}$ & 0.060$^{+0.098}_{-0.060}$ &
0.001$^{+0.067}_{-0.001}$ \\

& Norm ($\times$ $10^{-5}$) & 1.15$^{+0.33}_{-0.33}$ & 0.92$^{+0.33}_{-0.33}$
& 0.96$^{+0.30}_{-0.30}$ & 0.88$^{+0.35}_{-0.35}$ & 1.24$^{+0.31}_{-0.31}$ \\
[1ex]

$\chi^2_{\nu}$ (d.o.f.) & & 0.90 (69) & 0.80 (83) & 0.82 (76) & 1.08 (75) & 1.25 (75) \\
[1ex]

\hline \hline

\multicolumn{1}{l}{Model} & \multicolumn{1}{l}{Component}&
\multicolumn{1}{r}{0.6} & \multicolumn{1}{r}{0.7} &
\multicolumn{1}{r}{0.8} & \multicolumn{1}{r}{0.9} &
\multicolumn{1}{r}{1.0} \\

\hline

wabs & $N_{\rm{H}}$ ($\times$ $10^{22}$ cm$^{-2}$) &
0.025$^{+0.019}_{-0.015}$ & 0.021$^{+0.019}_{-0.016}$ & 0.145$^{+0.024}_{-0.020}$ & 0.052$^{+0.014}_{-0.012}$ &
0.079$^{+0.030}_{-0.023}$ \\ [1ex]

pcfabs1 & $N_{\rm{H}}$ ($\times$ $10^{22}$ cm$^{-2}$) &
19.7$^{+1.1}_{-1.1}$ & 21.7$^{+1.3}_{-1.2}$ &
16.5$^{+1.1}_{-1.0}$ & 26.5$^{+1.3}_{-1.3}$ & 26.2$^{+1.6}_{-1.5}$
\\ [1ex]

& CoverFrac & 0.793$^{+0.013}_{-0.013}$ & 0.731$^{+0.013}_{-0.013}$ &
0.697$^{+0.013}_{-0.013}$ &
0.776$^{+0.010}_{-0.010}$ & 0.828$^{+0.016}_{-0.016}$
\\ [1ex]

pcfabs2 & $N_{\rm{H}}$ ($\times$ $10^{22}$ cm$^{-2}$) &
6.32$^{+0.36}_{-0.33}$ &
5.78$^{+0.28}_{-0.26}$ & 3.75$^{+0.16}_{-0.15}$ & 6.37$^{+0.27}_{-0.25}$ &
11.26$^{+0.79}_{-0.74}$
\\ [1ex]

& CoverFrac & 0.980$^{+0.003}_{-0.002}$ & 0.980$^{+0.003}_{-0.002}$ & 0.980$^{+0.003}_{-0.003}$ &
0.980$^{+0.001}_{-0.002}$ & 0.980$^{+0.003}_{-0.004}$
\\ [1ex]

MEKAL1 & kT & 0.180$^{+0.022}_{-0.022}$ & 0.199$^{+0.024}_{-0.023}$ & 0.147$^{+0.013}_{-0.013}$ &
0.128$^{+0.009}_{-0.009}$ & 0.160$^{+0.023}_{-0.023}$ \\

& Norm ($\times$ $10^{-3}$) &
3.65$^{+0.75}_{-0.75}$ & 3.68$^{+0.78}_{-0.78}$
& 33.06$^{+7.90}_{-7.90}$ & 5.72$^{+1.00}_{-1.00}$ & 3.17$^{+0.98}_{-0.98}$ \\
[1ex]

MEKAL2 & kT & 3.01$^{+0.67}_{-0.48}$ & 2.73$^{+0.90}_{-0.67}$ & 1.71$^{+0.37}_{-0.27}$ &
6.35$^{+4.46}_{-3.34}$ & 1.18$^{+0.34}_{-0.19}$ \\

& Norm ($\times$ $10^{-5}$) &
3.05$^{+0.78}_{-0.78}$ & 2.80$^{+1.15}_{-1.15}$ & 4.00$^{+1.07}_{-1.07}$
& 2.25$^{+1.02}_{-1.02}$ & 2.06$^{+1.16}_{-1.16}$ \\
[1ex]

MEKAL3 & kT & 40.8$^{+12.5}_{-9.6}$ & 28.6$^{+10.2}_{-6.8}$ & 30.6$^{+8.0}_{-6.6}$ &
14.8$^{+1.9}_{-1.2}$ & 32.6$^{+17.8}_{-10.5}$ \\

& Norm ($\times$ $10^{-3}$) & 5.59$^{+0.14}_{-0.14}$ & 6.48$^{+0.15}_{-0.15}$ & 5.42$^{+0.12}_{-0.12}$ &
7.30$^{+0.16}_{-0.16}$ & 5.55$^{+0.18}_{-0.18}$  \\

Gaussian & line centre (keV) &
6.31$^{+0.05}_{-0.06}$ & 6.36$^{+0.03}_{-0.03}$ & 6.36$^{+0.04}_{-0.04}$ & 6.36$^{+0.03}_{-0.04}$ &
6.42$^{+0.03}_{-0.03}$ \\

& $\sigma$ &
0.086$^{+0.069}_{-0.061}$ & 0.075$^{+0.047}_{-0.042}$ & 0.114$^{+0.053}_{-0.038}$ & 0.053$^{+0.054}_{-0.053}$ &
0.036$^{+0.041}_{-0.036}$ \\

& Norm ($\times$ $10^{-5}$) &
1.09$^{+0.34}_{-0.34}$ & 1.79$^{+0.37}_{-0.37}$ & 1.76$^{+0.37}_{-0.37}$ & 1.08$^{+0.33}_{-0.33}$ & 1.07$^{+0.32}_{-0.32}$  \\
[1ex]

$\chi^2_{\nu}$ (d.o.f.) & & 0.83 (73) & 1.07 (80) & 0.73 (83) & 0.74 (66) & 0.82 (72) \\
[1ex]

\hline
\hline
\end{tabular}
%\end{center}
%\end{minipage}
\end{table*}
\vspace{0.3cm}

In order to investigate the changes of this composite model spectrum
over the orbital phase, we performed orbital phase-resolved
spectroscopy. We extracted spectra for each 0.1 orbital phase
interval, and fitted them with the composite model described above.
Table 1 displays the spectral parameters over the orbital phases.
Figure 4 shows the 10 phase resolved spectra extracted from the
data. The fits to the spectra clearly reveals that the neutral
Hydrogen column density of the second partial covering absorber {\sc
pcfabs2} increases up to 5 times during the the orbital dipping
phases (see Figure 5, and Table 1), while the other partial covering absorber ({\sc pcfabs1})
shows no particular variation (See Table 1). The first
two plasma temperatures ($kT1$ and $kT2$) show no significant
variation over the orbital phase, however the third temperature
($kT3$) peaks during the orbital minima indicating spectral
hardening (See Figure 5 and Table 1).

The shape of the spectrum shows distinct variations over the orbital
phase as well. During the phases where X-ray flux is minimum (i.e.
phases around 0-0.1) there is a substantial decrease in normalized
count rates around 1-2 keV regime (see Figure 4), while below 1 keV
the normalized count rates stay at the same level with that of the
rest of the phases. This soft excess brought our attention to a
possible warm absorber in the line of sight. In order to investigate
this feature we have extracted two spectra from orbital maximum and
minimum regions (phases between 0.55-0.85 and 0.9-1.2 respectively),
and fitted them with the composite model similar to the one in Table
1, this time replacing the second partial covering absorber and one
of the plasma emission components with a warm absorber model (i.e.
{\sc warmabs} model implemented into {\sc xspec}). {\sc warmabs}
models the absorption from a photoionized plasma in the line of
sight with column densities of ions (including small cross sections)
coupled through a photoionization model using stored level
populations calculated by XSTAR (Kallman \& Bautista 2001) assuming
a given continuum spectrum. The  fit to the orbital minima and
maxima yielded results with $\chi^2_{\nu}$ of 1.39 and 1.24; with values of
$N_{\rm{H}}$ = 2.09$^{+0.98}_{-1.09}$$10^{22}$ cm$^{-2}$ and 0.56$^{+0.26}_{-0.15}$
$\times$$10^{22}$ cm$^{-2}$; and log($\xi$) =
0.23$^{+0.37}_{-0.26}$ and $<$0.30 erg cm s$^{-1}$,
respectively. Here $N_{\rm{H}}$ is the equivalent hydrogen column
density of ionized absorption and $\xi = L /n_{\rm e} ~ r^{2}$ is
the ionization parameter where L is the luminosity of the ionizing
source, $n{\rm _e}$ the electron density of the plasma and r the
distance between the absorber and the ionizing source. The fitted
spectra of the orbital minima and maxima are presented in Figure 6,
while the fitted parameters are given in Table 2.

\begin{table}[ht!]
\centering
%\begin{minipage}{140mm}
\label{1} \scriptsize{ \caption{Spectral parameters derived from the
fits to the orbital maxima and minima. The model used is {\sc
warmabs*wabs*pcfabs*(mekal+mekal+gauss)}. The gaussian emission line
energy is fixed at 6.4 keV. $N_{\rm{H}}$ of the {\sc warmabs} model is
the equivalent hydrogen column density of ionized warm absorption. All the errors are given in 99\%
confidence level.} }

%\begin{center}
\begin{tabular}{@{}llrrrrrrrrrrr@{}}
\hline \hline

\multicolumn{1}{l}{Model} & \multicolumn{1}{l}{Component}&
\multicolumn{1}{r}{Maxima} & \multicolumn{1}{r}{Minima} \\

\hline

warmabs & $N_{\rm{H}}$ ($\times$ $10^{22}$ cm$^{-2}$) &
0.56$^{+0.26}_{-0.15}$ & 2.09$^{+0.98}_{-1.09}$ & \\

& log($\xi$) & $<$ 0.30  & 0.23$^{+0.37}_{-0.26}$
\\ [1ex]

& vturb (km s$^{-1}$) & 295$^{+295}_{-295}$ & 193$^{+108}_{-193}$
\\ [1ex]

wabs & $N_{\rm{H}}$ ($\times$ $10^{22}$ cm$^{-2}$) &
0.25$^{+0.12}_{-0.08}$ & 0.25$^{+0.26}_{-0.24}$
\\ [1ex]

pcfabs & $N_{\rm{H}}$ ($\times$ $10^{22}$ cm$^{-2}$) &
10.9$^{+1.2}_{-1.0}$ & 20.7$^{+1.9}_{-1.8}$
\\ [1ex]

& CoverFrac & 0.88$^{+0.02}_{-0.02}$ & 0.98$^{+0.01}_{-0.01}$
\\ [1ex]

MEKAL1 & kT & 0.081$^{+0.038}_{-0.081}$ & 0.19$^{+0.11}_{-0.08}$
\\

& Norm & 0.058$^{+0.121}_{-0.051}$ & 0.18$^{+4.54}_{-0.13}$ \\
[1ex]

MEKAL2 & kT & 61.1$<$ & 49.5$<$ \\

& Norm  & 0.014$^{+0.001}_{-0.001}$ &
0.012$^{+0.01}_{-0.01}$ \\
[1ex]

Gaussian & $\sigma$ & 0.23$^{+0.06}_{-0.05}$ &
0.16$^{+0.12}_{-0.06}$ \\

& Norm ($\times$ $10^{-5}$) & 8.4$^{+1.7}_{-1.6}$ &
5.0$^{+2.0}_{-1.8}$ \\
[1ex]

$\chi^2_{\nu}$ (d.o.f.)  & & 1.24 (216) & 1.39 (126) \\
[1ex]

\hline \hline

\end{tabular}
\end{table}

\begin{figure}
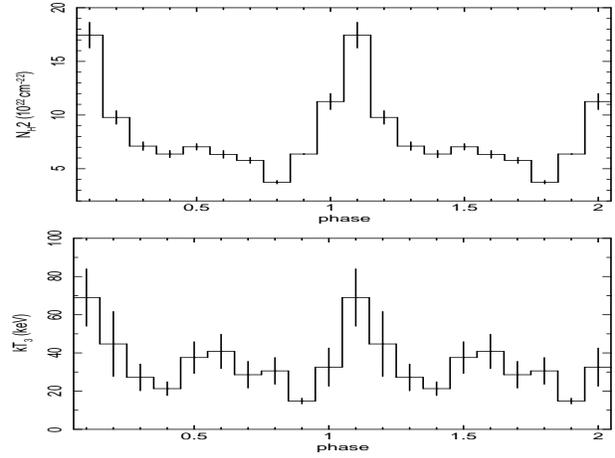

\centerline{
\includegraphics[width=3cm,height=8cm,angle=270]{nh2_updated.ps}}
\centerline{
\includegraphics[width=3cm,height=8cm,angle=270]{kt3_updated.ps}}

\caption{The variation of $N_{\rm{H}}$ parameter of the partial
covering absorber {\sc pcfabs2} on top and
the variation of the temperature parameter of the hardest emission component
{\sc mekal3} at the bottom. Both variations are over the orbital phase.}
\end{figure}

\section{Discussion}

The orbital modulation in CVs provides essential information on the
structure of the accretion disk in non-magnetic systems and
intermediate polars. The orbital-phase resolved analysis requires
demanding conditions in the X-ray regime such as adequate count
rates, sufficient observation times and suitable orbital/spin
periods. Previous works on orbital phase resolved analysis (e.g.,
Pek\"on \& Balman 2011, Hellier et al. 1996, Itoh et al. 2006,
Salinas \& Sclegel 2004, Staude et al, 2008) showed that the
spectral characteristics such as absorption column densities, plasma
temperature distributions, line widths from IPs can be studied in
detail over the orbital phase which contributes to the understanding
of the detailed structure of the accretion phenomenon of these
systems. In this work, we tried to incorporate the orbital-phase
resolved spectroscopy in order to calculate the absorption
explicitly from the disk and the accretion curtains/columns,
separately. The short P$_{spin}$ of the system is an advantage since
the spectral contributions from the spin modulation can average out in a
single orbital phase bin.

\begin{figure*}[ht!]
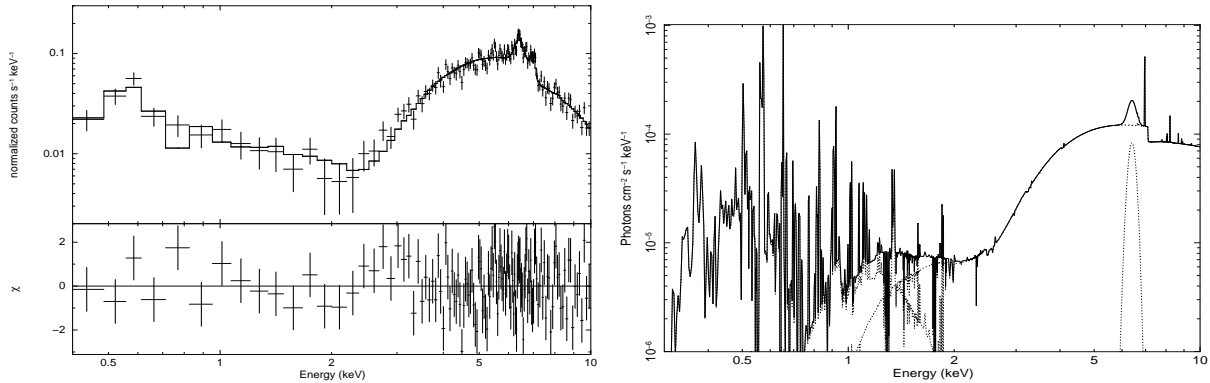

\centerline{
\includegraphics[width=5cm,height=8cm,angle=270]{warm_min_n08_4.ps}
\includegraphics[width=5cm,height=8cm,angle=270]{warm_min_modelplot.ps}}

\caption{The composite spectral fit including a warm absorber
fitted to the spectrum at orbital minima of the source (phases
between 0.9 and 1.2) is shown on the left. On the right, a plot of the composite model
alone without the data is presented.}
\end{figure*}

It was pointed out by Evans et al. (2004) that the orbital
modulation of FO Aqr may be a result of the absorption from
vertical structures on the disk extending out to 140$^\circ$ on the
plane of the accretion disk and 25$^\circ$ above the accretion plane
(however not explicitly calculated). Moreover; Parker, Norton \&
Mukai (2005) showed that the modulation depth increases at lower
energies which supports the absorption effects from the disk. The folded
light curve over the orbital period shows three features: a plateau
feature with almost constant/slighly increasing flux between phases
0.3-0.6; a peak at phases 0.6-0.8, and a quick decline and deep
feature at the rest of the phases where flux minimizes around phase
0-0.1. In our orbital phase resolved analysis, absorption column of
the second partial absorber ({\sc pcfabs2}) follows this trend
inversely, where it peaks during the orbital dipping phases and is
lower during the orbital peak as shown in Figure 5. The
absorption column density of the maximum and minimum phases are
$N_{\rm{H}}$ value of 17.45$\times$$10^{22}$ cm$^{-2}$ at the
maximum reaching up to 5 times the minimum value of
3.75$\times$$10^{22}$ cm$^{-2}$. The covering fraction of this absorber shows no
significant modulation over the orbital phase and is close to 1.
Hence the absorber is almost uniformly obscuring the X-rays in the
line of sight. The absorbing material is well spread out on the disk
since absorbing column never vanishes over the orbital period, but
is denser and colder at dipping phases causing more absorption (See
Figure 7).

\begin{figure}[ht!]
\centerline{
\includegraphics[width=8cm,height=8cm,]{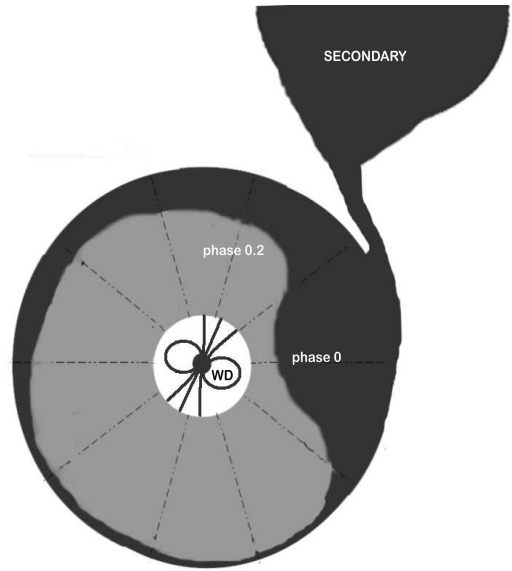}}

\caption{The schematic diagram showing the system. The absorber
(represented as the dark region on the disc) is concentrated around
the accretion impact zone (phase 0), and smeared out gradually
around the accretion disk.}
\end{figure}

Osborne \& Mukai (1989) suggest that the optical eclipse occurs at
the inferior conjunction of the secondary star. Preceding the
optical eclipse on the binary plane by 0.2 orbital phase is a
reprocessing site noted by these authors, which is most likely the
accretion stream-accretion disk interaction zone. When we phase-lock
the X-ray light curve to the optical light curve using the
ephemerides given by Osborne \& Mukai (1989) HJD 2446081.3037 +
0.20205956E for the optical eclipse, we find that the X-ray orbital
minimum is at phase 0.8. Our phase 1.0 or 0.0 is phase-locked to the
UV minimum given by Evans et al. (2004). This is the same phase 0.8
in the optical band.

The absorption column for the simple photoelectric absorber
component ({\sc wabs}), $N_{\rm{H}}$, has values around
0.1$\times$$10^{22}$ cm$^{-2}$ and the first partial covering absorber
component has an $N_{\rm{H}}$ value around 20$\times$$10^{22}$
cm$^{-2}$ and the covering fraction is around 0.6-0.8 . Considering
the variation of these three absorption components over the orbital
phase (see Table 1), it can be inferred that each has different
origins. The simple absorber with relatively lower column density
than the others (i.e. {\sc wabs}) account for the interstellar
absorption. The first partial covering absorber ({\sc pcfabs1}) with
changing covering fraction and relatively high absorbing column
accounts for the absorption arising from the accretion
curtain/column since it does not vary with the orbital phase. And
finally the second partial covering absorber ({\sc pcfabs2})
accounts for the absorption from the bulk material on the disk
(edge) since it changes with the orbital motion as explained above.
At any orbital phase in our analysis, we can see the combined
effects of these three absorption components. The interstellar
component is in effect at all phase intervals since it is always in
the line of sight; and the absorption from the accretion
curtain/column is averaged out in each orbital-phase bin since a
phase bin of 0.1 spans an interval of 1745.8 s which is greater than
the spin period of the white dwarf.

The behavior of the spin pulse profile over the orbital period of
the system also supports this scenario of the absorption components.
The spin pulse shape have a sinusoidal nature that is constant
throughout the orbital motion. The amplitude of the spin pulse vary
in line with the orbital folded light curve. Therefore the
variations over the spin and orbital periods have different origin,
former by the absorption from the accretion curtains/columns and
latter by absorption from the material on the disk near the
accretion stream impact zone.

The average spectrum of the source can be represented with 3 plasma
emission components; a soft component around 0.15 keV, a medium
component around 3 keV and a hard component around 30 keV. The hard
component temperatures obtained from the orbital phase resolved
spectroscopy ranges between 15-70 keV. The phase averaged XMM-Newton
data yields 13.9$^{+6.5}_{-3.1}$ keV (Evans et al. 2004), $Suzaku$
data yields 14$^{+4.0}_{-1.4}$ keV (Yuasa et al. 2010) and
$Swift$-$XRT$ and $INTEGRAL$-$IBIS$ data yields
29.7$^{+70.1}_{-16.6}$ keV (Landi et al. 2009) which are all
consistent with this range. The first two softer plasma temperatures
do not follow a significant variation over the orbital phase. However,
during the orbital maximum at around phase 0.8, the flux of
the soft component peak at about 10 times more than the average. On the other
hand, the temperature of the hard component peaks during the orbital
dipping phases. These characteristics indicate a signature of
spectral softening during the orbital maximum and hardening during
the orbital minimum.

Hellier et al. (1993) and Parker, Norton \& Mukai (2005) argued that
the orbital modulation in IPs may be similar to those of Low Mass
X-ray Binaries (LMXBs) where non-symmetric structures on the
accretion disk raised by the impact stream (possibly around the hot
spot) causing energy dependent absorption hence reducing the X-ray
flux. In particular a class of LMXBs called Low-mass X-ray binary
dippers show considerable fluctuations and modulation around the
orbital minimum. Such characteristics have been explained by
modeling the dip and non-dip spectra using highly ionized warm
absorber models (Boirin et al. 2005, Diaz-Trigo et al. 2006, Balman
2009 and the references therein). We tried to explore the warm
absorber scenario for FO Aqr extracting spectra from phases around
the orbital minima and maxima and fit them with the assumed
composite spectral model including a warm absorber. We simply
replaced the absorption component due to the disk ({\sc pcfabs2})
and one of the {\sc mekal} components with {\sc warmabs} model. Both
minima and the maxima could successfully be modeled with an ionized
absorber. The ionized absoption N$_{\rm{warmabs}}$ value increases
by 4 times during the orbital minima which is expected since the
absorber is denser and colder at the orbital "dip". The values for
the ionization parameter log($\xi$) is consistent within error
margins between the minima and the maxima. This may be due to the
fact that we used a range of 0.3 phases for the minimum and maximum
calculations. The N$_{\rm{warmabs}}$ values we find are lower than
those found in LMXBs by about 10 times at all times (min or max)
(see Diaz-Trigo et al. 2006). Since $\xi$ is given by L/(n$_{\rm e}$
$r^{2}$), $\xi$ values can be compared using $\xi$ =
L/N$_{\rm{H}}$R$_{\rm d}$ where n$_{\rm e}$ $\sim$
N$_{\rm{H}}$/R$_{\rm d}$ (R$_{\rm d}$, disk radius and N$_{\rm{H}}$
equivalent hydrogen column density of ionized absorption). The
0.6-10 keV luminosities of LMXB dippers listed by Diaz-Trigo et al.
(2006) are of the order of 10$^{36}$ erg s$^{-1}$. The X-ray flux of
FO Aqr in 0.6-10 keV range is 1.4$\times$$10^{-10}$ ergs cm$^{-2}$
s$^{-1}$, yielding a luminosity of the order of 10$^{32}$ erg
s$^{-1}$ at 400 pc source distance. Therefore, one can approximate
log($\xi$$_{\rm CV}$)$\sim$ -3 log($\xi$$_{\rm LMXB}$) (given the
warmabs absorption column is a factor of 10 more in LMXB dippers).
We also assumed similar $P_{orb}$ and R$_{\rm d}$, disk radius. This
yields a range of log($\xi$$_{\rm CV}$)$\sim$ -0.8--1. Therefore the
orbital dip  in the X-ray light curve of FO Aqr is analogous to the
dips observed in the dipping LMXBs with the expected cause of the
dips being variation of temperature and density of the warm ionized
material on the disk. A schematic representation of the system is
shown in Figure 7.

\section{Conclusions}

We have presented X-ray orbital-phase resolved analysis of the
intermediate polar FO Aqr. This work improves upon the previous
works by calculating the spectral parameters over the orbital phase.
The distinction between the absorbing components are clarified and
the values are explicitly calculated. The absorption originating
from the polar regions of the white dwarf can be resolved from the
absorption by structures on the accretion disk. The shape of the
spin pulse profile is unaffected by the orbital motion, however the
semi-amplitude of the profile change over the orbital phase. The
X-ray orbital variation over the orbit in the system arises from
absorption by the bulge material on the disk spread well over the
disk. The absorption column is greatly enhanced during the X-ray
orbital dip. Moreover, we have modeled the absorption from the
orbital dip with a warm absorber model for the first
time for this source and also for CVs, confirming the ionized nature of the material
on the disk causing the absorption. We derived a range of ionization
parameter log($\xi$$_{CV}$)$\sim$
-0.8--1 for plausible warm absorbers on CV disks. We calculate that the orbital
modulations in FO Aqr (and plausibly some other CVs) are similar to those seen in LMXB dippers as
variations (density/temperature) of a warm absorber at the accretion
impact region and spreading on the disk.

\section*{Acknowledgments}

Authors thank D. de Martino and K. Mukai for valuable discussions.
The authors acknowledge support from T\"UB\.ITAK, The Scientific and
Technological Research Council of Turkey, through project 108T735.

\label{lastpage}

\end{document}